\documentclass[twocolumn,showpacs]{revtex4}
\usepackage{rotating,graphicx,dcolumn}     

\begin{document} 
\preprint{Version 31.01.2001}

\title{Far-Infrared Excitations below the Kohn Mode:\\
       Internal Motion in a Quantum Dot}
\author{Roman Krahne, Vidar Gudmundsson$^*$, 
        Christian Heyn, and Detlef Heitmann}
\affiliation{Institut f\"ur Angewandte Physik und Zentrum
             f\"ur Mikrostrukturforschung,\\ Universit\"at Hamburg,
             Jungiusstra\ss e 11, D--20355 Hamburg, Germany\\
             $^*$Science Institute, University of Iceland,
             Dunhaga 3, IS-107 Reykjavik, Iceland}

%

\begin{abstract}
We have investigated the far-infrared response of  quantum dots in
modulation doped GaAs heterostructures. We observe novel modes at
frequencies below the center-of-mass Kohn mode. Comparison with
Hartree-RPA calculations show that these modes arise from the
flattened potential in our field-effect confined quantum dots.
They reflect pronounced relative motion of the charge density with
respect to the center-of-mass.

\end{abstract}


\maketitle

According to the generalized Kohn's theorem the far-infrared (FIR)
response of a quantum dot with parabolic external potential in a
magnetic field $B$ consists, independent of the number of
electrons in the quantum dot, of only two modes with
dispersion:\cite{Maksym90:108}
\begin{equation}
      \omega_{\pm}=\sqrt{\Omega_0^2+(\omega_c/2)^2}\pm\omega_c/2.
\label {dot}
\end{equation}
Here $\Omega_0$ describes the external potential,
$\omega_c=eB/m^*$ is the cyclotron frequency, $m^*$ the effective
mass. One mode, $\omega_+$, increases with increasing magnetic
field and approaches the cyclotron frequency $\omega_c$, the other
mode, $\omega_-$, decreases in frequency with decreasing magnetic
field. Both modes represent rigid center-of-mass motion of all
electrons in the dot. This behavior is actually very often
observed in experiments on quantum dots which are prepared from
semi\-conductor hetero\-structures.\cite{sikorski89.1,meurer92.2}
The reason being the electrostatic environment causing the external
potential to have a nearly perfect parabolic shape. In other
experiments, in particular on etched quantum dots with a larger
number of electrons and a more hard-wall type of external
potential, one observes additional sets of modes at {\it higher}
frequencies (see for example Ref.\onlinecite{demel90.1}).
Such a mode spectrum
can be calculated using a mean field approach and random phase
approximation (RPA).\cite{Gudmundsson91:12098} They can be
visualized in terms of `confined' plasmon modes, for example in
the model of Fetter \cite{fetter86.1} (See also below for
details).

We have performed experiments on field-effect-confined quantum
dots in AlGaAs/GaAs-hetero\-structures. We find in FIR experiments
additional modes {\it below} the high-frequency Kohn mode, $\omega_+$,
which are, however, definitely higher than $\omega_c$. The
cyclotron resonance itself is not observed for these isolated
dots. We compare our experiments with self-consistent Hartree
calculations. From this comparison we find that the new modes
arise from the flattened external potential in our
field-effect-induced quantum-dot array. Our calculation shows that
this new mode represents a dipole active charge-density
oscillation involving strong relative motion of the electrons with
respect to the center-of-mass.

Quantum-dot arrays were prepared from modulation-doped
Al$_{0.33}$Ga$_{0.67}$Al-GaAs heterostructures with a
heterostructure interface-to-surface distance of 55 nm. A
Si-$\delta$ doped layer was grown 300 nm underneath the
AlGaAs-GaAs interface and served as a backgate to charge the
quantum dots. The sample design is sketched in the inset of
Fig.~1. The electron density and mobility at 1.8 K were
$N_S=3.8\cdot 10^{11}$ cm$^{-2}$ and $\mu=300\,000$ cm$^2$/Vs,
respectively. A photoresist-dot array was prepared by holographic
lithography onto the sample surface with a period of $a = 330$ nm,
the diameter of the photoresist dots was 150 nm and the height
about 170 nm. A 7 nm semitransparent Ti gate with 3 mm diameter
was evaporated onto the photoresist dot array. The gate voltage
was applied between the topgate and the $\delta$-doped backgate.
The latter was contacted outside the active mesa where the
AlGaAs-GaAs was removed by wet etching. The experiments were
performed in a superconducting magnet cryostat connected by wave
guides to a Fourier transform spectrometer. We show in the
following the normalized transmission $T(V_G)/T(V_T)$, where $V_T$
is the threshold value at which the electron system is totally
depleted. The spectral resolution was set to 1 cm$^{-1}$. The
temperature was 1.8 K.

Experimental spectra for a dot array with $N$ = 30 electrons per
dot are shown in Fig.~1. This electron number $N$ can be
determined from the known oscillator strength at high magnetic
fields ($B=8$ T in our case), see for example
Refs.\onlinecite{sikorski89.1} and \onlinecite{meurer92.2}.
Spectra taken at small magnetic fields are displayed in Fig.~1(a).
For $B=0$ T one resonance is observed which splits into two modes
with increasing magnetic field. One mode increases with increasing
$B$, the other decreases. This behavior can be well understood by
the rigid center-of-mass motion of the electrons in the dot
described in the introduction. In Fig.~1(b) we find for higher
magnetic fields clearly an additional resonance below the dominant
$\omega_+$ mode. At smaller $B$ in the regime from 1.5 T to 3.5 T
this additional feature at the low-energy side of the $\omega_+$
mode is very broad and might consist of one or more resonances. At
higher magnetic fields $B\ge 4.5$ T clearly a double-peak
structure is observed, i.e., one mode in addition to the
$\omega_+$ mode. For the lower mode we notice that it increases
faster in frequency than the higher mode (i.e.,\ they are not
parallel) and that its resonance half width decreases with $B$.
Figure\ 1(c) shows spectra of the same sample at a gate voltage where
we have 6 electrons per dot. Here again we observe, at high
magnetic fields, a distinct double-peak structure where the
energetically lower mode increases more rapidly in frequency with
increasing $B$. Since the behavior of the additional resonance is
qualitatively the same for 30 and for 6 electrons we conclude that
it is an intrinsic property of the electron system for our
potential that does not depend significantly, at least at larger
$B$, on the electron number.

The magnetic-field dispersions extracted from the spectra are
plotted in Fig.\ 2. We have fitted the lower branch and the
sharper high-frequency branch with the dispersion expected from
the Kohn theorem for a parabolic external potential according to
eq.\ (1). From this equation we can also determine $\omega_c$. The
cyclotron resonance itself is, as expected for isolated dots, not
observed. The new mode has a frequency, which is definitely larger
than $\omega_c$. The observation of a new mode, in addition to the
Kohn modes, implies that the external potential is not parabolic.
Actually, since the external potential is formed by the field
effect of our modulated gate, we expect that it should be
flattened at higher energies were it eventually overlaps with the
neighboring dots of the array.

To confirm this explanation and to get a deeper microscopic
insight into this new type of excitation we have performed
self-consistent Hartree calculations for differently shaped
potentials and calculated the dynamic response within the RPA. The
details of the applied formalism have been described in a previous
publication.\cite{gudmundsson95.1} We have modeled an external
potential, with a soft bottom and a step  that flattens  for
higher energy, by the expression
\begin{equation}
      V(x)=ax^2+bx^4+W(x),
\end{equation}
where $x=r/a_0^*$ is the radial coordinate scaled by the
effective Bohr radius $a_0^*=9.77$ nm in GaAs and
\begin{equation}
      W(x)=c\left[1-f\left( 3.9x-12 \right)\right],
\end{equation}
with $f(x)=1/(\exp(x)+1)$. This potential for $a=0.48$ meV,
$b=-1.8\times 10^{-3}$ meV, and $c=6$ meV will be shown later in
Fig.~5. The calculated FIR absorption is plotted in a 3D plot in
Fig.~3. The dispersions of modes with an oscillator strength
larger than $10^{-18}$ on the scale of Fig.~3 is plotted in Fig.~4. We
see from these figures that the absorption is dominated by the
strong Kohn modes.

In the regime between 1 to 2 T we have obtained in the calculation
a very complex mode spectrum. In this regime interaction with
Bernstein modes occurs which has been extensively discussed
previously.\cite{gudmundsson95.1} We will not elaborate on this
here. We note that indeed also the experimental dispersion, see
Fig.~2(b) at small $B$, shows a complex mode spectrum which seems to
be very sensitive on the electron number, in particular whether or
not we have partially or completely filled shells, and on the
shape of the potential. So it is very hard to develop a detailed
picture here. However, with increasing magnetic field the new mode
is clearly resolved both in the experiment and the calculation. To
get a microscopic insight into this new mode we have calculated in
Fig.~5 the equilibrium density and the induced density. The latter
is the density induced by the FIR dipole field in a certain moment
of time. The induced density for the $\omega_+$ and $\omega_-$
modes shows indeed a nearly perfect rigid displacement which
justifies that we call these modes still `Kohn modes'. However,
the new mode represents a complex charge oscillation with several
nodes. This is clearly an internal relative motion of the involved
electrons.

In many other experiments, in particular on quantum dots defined
in etched structures, also additional modes have been observed,
however, with frequencies {\it higher} than the $\omega_+$ Kohn
mode (see for example Ref.\onlinecite{demel90.1}). They  follow
approximately the dispersion given by the model of
Fetter\cite{fetter86.1}, who finds for a circular disk of a
two-dimensional electron system (2DES) with density $N_S$:

\begin{equation}
      \omega_{i\pm}=\sqrt{\Omega_{0i}^2+(\omega_c/2)^2}\pm\omega_c/2.
\end{equation}
with
\begin{equation}
      \Omega^{2}_{0i}=
      \frac{N_{S}e^{2}}{2\bar{\epsilon}\epsilon_{0} m ^{*}} \cdot
      \frac{i}{R}, \quad i = 1,2,3,\dots
\label{qdisk}
\end{equation}
where $R$ is the radius of the disk and $\bar{\epsilon}$ the
effective dielectric function of the surrounding media.

These modes might be visualized in the model of `confined'
plasmons. Actually, if one looks into the calculated absorption in
Fig.~3, one sees, very faintly, but clearly resolved on an
enlarged scale, modes which start at about 8 meV at $B=0$ and
split with increasing $B$, as expected from eq.~(4) and (5).
Obviously, both experiments and calculations show that such
excitations are very weak in potential with a flattened profile.
We have also calculated the induced density for these `confined'
plasmons, see Fig.~5(d). This mode of course also involves relative
electron motion. However, this motion is different from the one of
the new mode, with 9.27 meV in Fig.~5(b),  it has additional nodes
in the lower part of  the flanks.

We have calculated the absorption strengths for several
differently shaped potentials in order to understand under which
conditions pronounced low-frequency modes or high-frequency
`confined' plasmon modes occur. We have calculated the FIR
absorption for a potential $V(r)=ar^2+br^4$ with a large value of
$b=1.46\times 10^{-2}$ meV. We find that  such a potential
produces, except for coupling with Bernstein modes, nearly only
the two Kohn modes, no lower frequency mode and extremely weakly
the higher-frequency `confined' plasmon modes. It was already
pointed out by Ye and Zaremba \cite{Ye94:17217} that the
observation of only the Kohn modes is not a good test to rule out
$r^4$ terms in the potential. We have also performed calculations
for our potential with a steeper step, $c$ = 18 meV in eq.(3). For
5 and 6 electrons we find stronger high frequency modes, but  also
very pronounced below-Kohn modes. From the experimental experience
it seems to be necessary to have a quantum dot with a larger
number of electrons, which usually requires also a dot of larger
size, to observe the higher frequency `confined' plasmons. One
reason of course is that, since the intensity of these modes is
very weak, for example 0.03 of the Kohn mode in
Ref.\onlinecite{demel90.1}, one needs a large number of electrons
to have enough signal strength to observe these resonances. Under
these conditions one has, at least so far, not observed the
below-Kohn mode. Perhaps, the large number of electrons has then
changed the electrostatic environment of the dots in a way that
does not allow the formation of these modes.

Finally we would like to discuss an additional interesting finding
for the new mode. In the upper panel of Fig.~6 we show the
resonance positions extracted from spectra taken at a fixed
magnetic field $B$ = 8 T for different gate voltages. In the lower
panel we have plotted the relative absorption strength,
$S^i/(S^1+S^2)$, $i$~=~1,2, where $S^1$ and $S^2$ are the
integrated experimental absorption strengths of, respectively, the
upper and lower mode. At $V_G$ = 0 we have a continuous 2DES and
observe the cyclotron resonance at the frequency $\omega_c$ and
additionally a magnetoplasmon resonance at a slightly higher
frequency. Also at $V_G$ = 0, we have  a small density modulation
of the 2DES with period $a$. Due to the grating-coupler effect of
this periodic modulation, 2D plasmons with wave vector $q = 2\pi /
a$ are excited. With decreasing gate voltage and correspondingly
decreasing averaged 2D density the plasmon frequency decreases. In
the regime from $V_G$ = -0.5 V to -0.25 V the dispersion
undergoes several anticrossing with Bernstein modes. At about
-0.55 V the dispersion increases again. Here isolated dots are
formed and the resonance develops into the Kohn mode of the
quantum dot. At the same time, the cyclotron resonance increases
in frequency and, interestingly, develops into the new below-Kohn
mode that we have discussed here. So in this interpretation one
could say that the below-Kohn mode has its origin in the cyclotron
resonance. However, one can interpret the disperion also in the
following way. The high-frequency 2D plasmon undergoes, with
decreasing gate voltage, an anticrossing with the increasing
cyclotron resonance which develops, in this picture, into the Kohn
mode. This picture is supported by the observation that the high
oscillator strength of the lower-frequency cyclotron resonance is
transferred to the high-frequency Kohn mode. Such an transfer of
oscillator strength is typical for an anticrossing behavior. 

Although, we want to be cautious in drawing a conclusive interpretation
concerning the character of the below-Kohn mode from the
interesting gate voltage dependence in Fig.~6,
there are other experimental findings supporting the anticrossing
hypothesis for the data in Fig 6:
At high magnetic fields the induced density of the new mode (Fig.\ 5b) is
reminiscent of that of the first confined plasmon in an
electron slab with hard boundaries;\cite{Dempsey92:1719}
The dipole moment originates essentially from the portion of the density
distribution in the center part of the dot just before the steeper
edges. The induced density that is relevant to the restoring
force is therefore sampling an average potential curvature of the dot
near the center. In contrast to this, the Kohn mode is sampling an
average potential curvature that includes the edges, leading to
a higher energy.
 
In summary, we have investigated field-effect-confined quantum-dot
arrays. We find new modes below the $\omega_+$ Kohn mode. From
comparison with Hartree-RPA calculations we can trace the origin
of this excitation back to the flattened potential in this
field-effect-confined quantum dots. We show that microscopically
this excitation involves a complex relative motion of the
electrons in the dot. Interestingly, related phenomenons have been
reported in the far-infrared absorption in a non-parabolic
one-dimensional quantum well wire with steep edge
boundaries.\cite{Wixforth94:215}
                                                   
We gratefully acknowledge support from the German Science
Foundation DFG through SFB 508 ``Quanten-Materialien'', the
Graduiertenkolleg ``Nanostrukturierte Festk{\"o}rper'', the
Research Fund of the University of Iceland, and the Icelandic
Natural Science Council.

\bibliographystyle{prsty}

\begin{thebibliography}{10}

\bibitem{Maksym90:108}
P.~A. Maksym and T. Chakraborty, Phys. Rev. Lett {\bf 65},  108  (1990).

\bibitem{sikorski89.1}
C. Sikorski and U. Merkt, Phys. Rev. Lett. {\bf 62},  2164  (1989).

\bibitem{meurer92.2}
B. Meurer, D. Heitmann, and K. Ploog, Phys. Rev. Lett. {\bf 68},  1371  (1992).

\bibitem{demel90.1}
T. Demel, D. Heitmann, P. Grambow, and K. Ploog, Phys. Rev. Lett. {\bf 64},
  788  (1990).

\bibitem{Gudmundsson91:12098}
V. Gudmundsson and R. Gerhardts, Phys. Rev. B {\bf 43},  12098  (1991).

\bibitem{fetter86.1}
A.~L. Fetter, Phys. Rev. B {\bf 33},  3717  (1986).

\bibitem{gudmundsson95.1}
V. Gudmundsson, Arne Brataas, Peter Grambow, Bernd Meurer, Thomas Kurth 
and Detlef Heitmann, Phys. Rev. B {\bf 51},  17744  (1995).

\bibitem{Ye94:17217}
Z.~L. Ye and E. Zaremba, Phys. Rev. B {\bf 50},  17217  (1994).
 
\bibitem{Dempsey92:1719}
J. Dempsey and B.~I. Halperin, Phys. Rev. B {\bf 45},  1719  (1992).
 
\bibitem{Wixforth94:215}
A. Wixforth et al., Semicond. Technol. {\bf 9}, 215 (1994).

\end{thebibliography}

\begin{figure}
\begin{center}
\includegraphics[bb=83 66 433 643,clip,width=8cm]{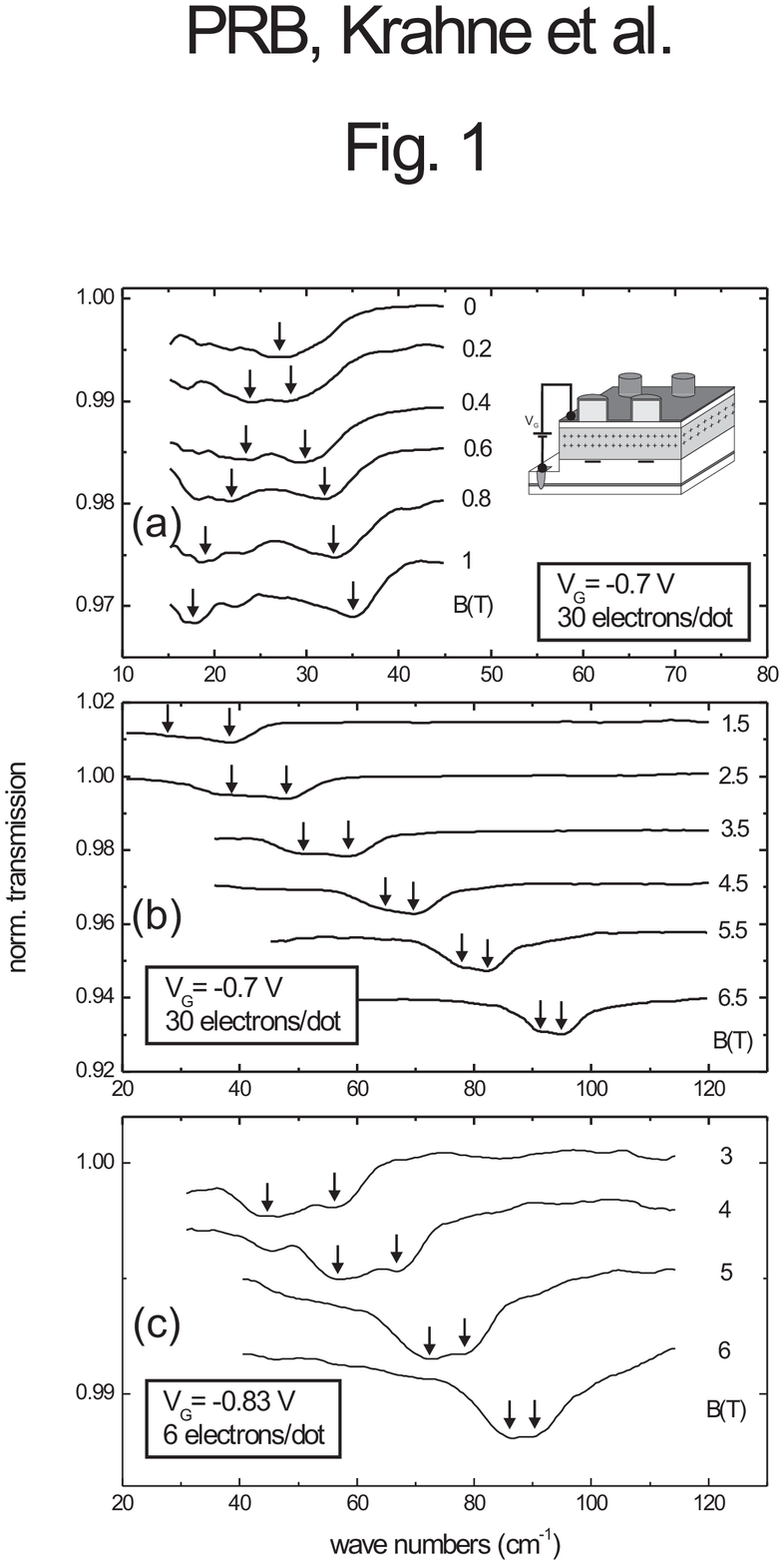}
\end{center}
\caption{Normalized transmission  spectra for different magnetic
         fields $B$. The spectra are shifted vertically for clarity. The
         arrows indicate the resonance frequency plotted in Fig.~2. At
         $V_G$ = -0.7 V in (a) and (b) we have 30 electrons per dot, at
         $V_G$ = -0.83 V in (c) 6 electrons. The inset sketches the design
         of  the sample.}
\label{fig1}
\end{figure}

\begin{figure}
\begin{center}
\includegraphics[bb=77 416 304 704,clip,width=8cm]{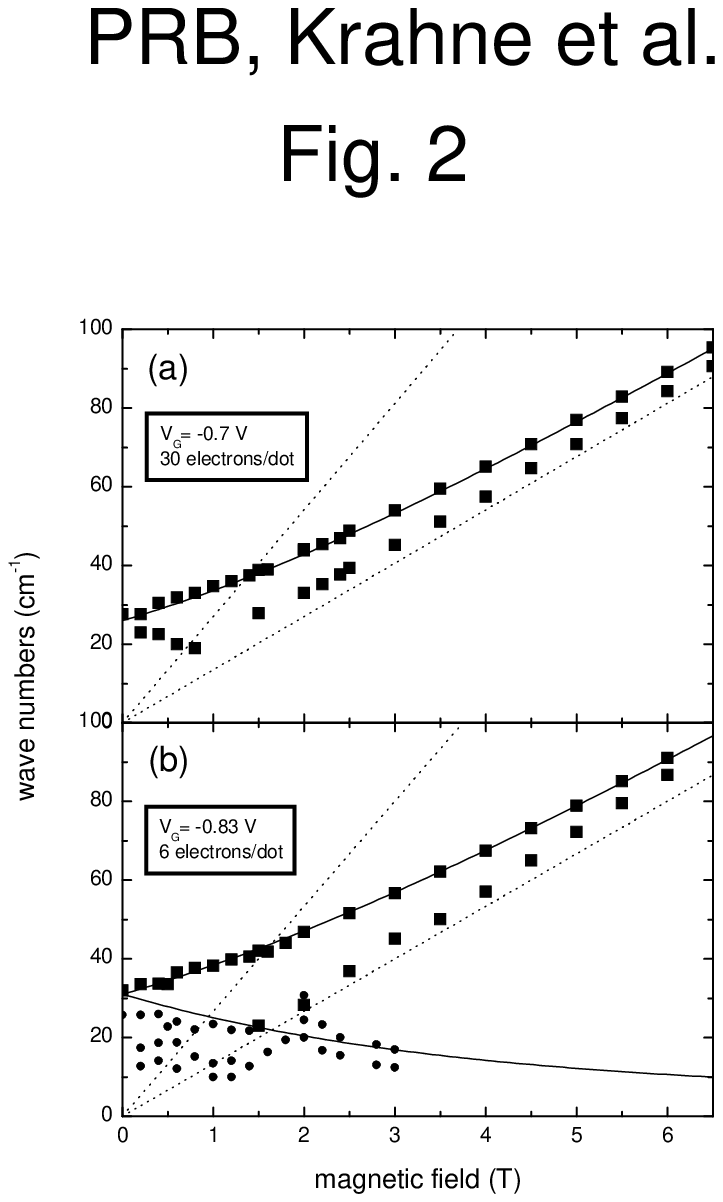}
\end{center}
\caption{Experimental dispersion for quantum dots with (a) 30
         electrons and (b) 6 electrons. Full lines are fits with the Kohn
         modes eq.(1), the dotted lines are $\omega_c$ and 2$\omega_c$
         extracted from this fit. A new mode, the below-Kohn mode, is
         observed {\it below} the upper Kohn mode but clearly {\it above}
         $\omega_c$.}
\label{fig2}
\end{figure}
\pagebreak

\begin{figure}
\begin{center}
\includegraphics[width=8cm]{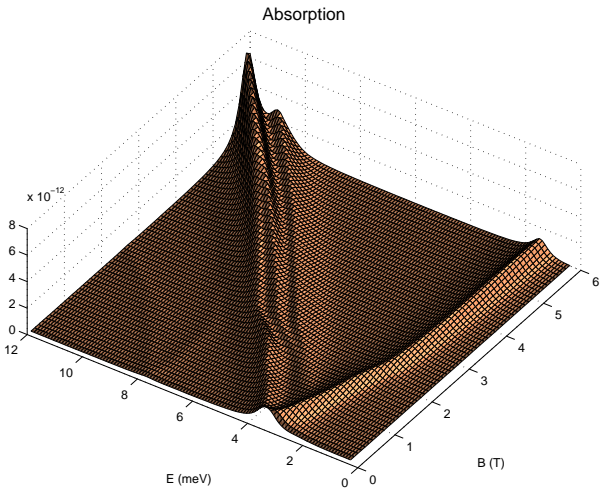}
\end{center}
\caption{Calculated dipole absorption for a quantum dot with 5 electrons
         in a flattened potential described in the text.
         In addition to the strong Kohn modes new
         modes below the high-frequency Kohn mode are found also in the
         calculation. The half-linewidth is 0.3 meV and $T=1$ K.}
\label{fig3}
\end{figure}

\begin{figure}
\begin{center}
\includegraphics[width=8cm]{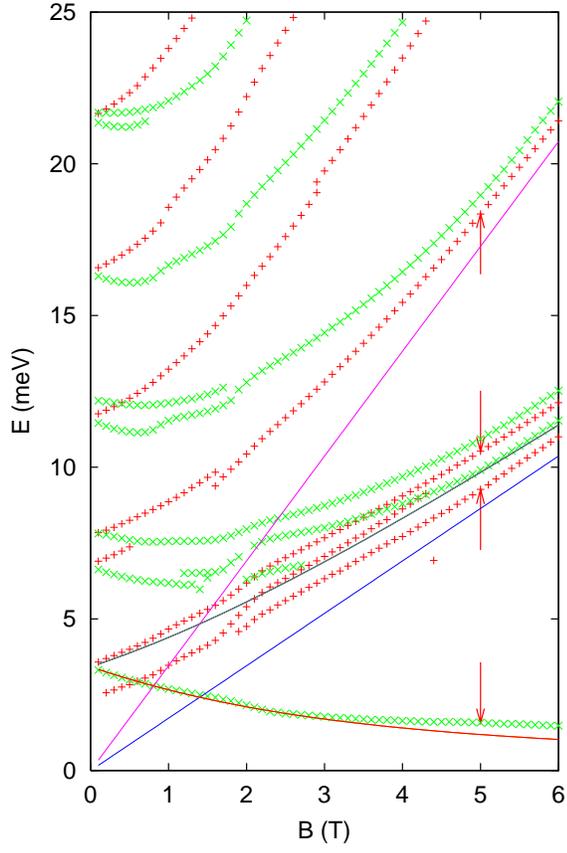}
\end{center}
\caption{Calculated dispersion for a quantum dot with 5 electrons
         and the flattened potential described in the text. Only resonances
         with an oscillator strength larger than $10^{-18}$ on the scale
         of Fig.~3 are plotted. For certain frequencies at $B$ = 5 T,
         marked by arrows, the induced density is calculated in Fig.~5.
         The half-linewidth is 0.3 meV and $T=1$ K. One circular
         polarization is marked by $+$ and the other by {\sf x}.
         The lines $E=\hbar\omega_c$ and $E=2\hbar\omega_c$ are indicated
         in the figure together with curves showing the Kohn modes
         for $\hbar\Omega_0=3.4$ meV.}
\label{fig4}
\end{figure}

\begin{figure}
\begin{center}
\includegraphics[width=6.5cm]{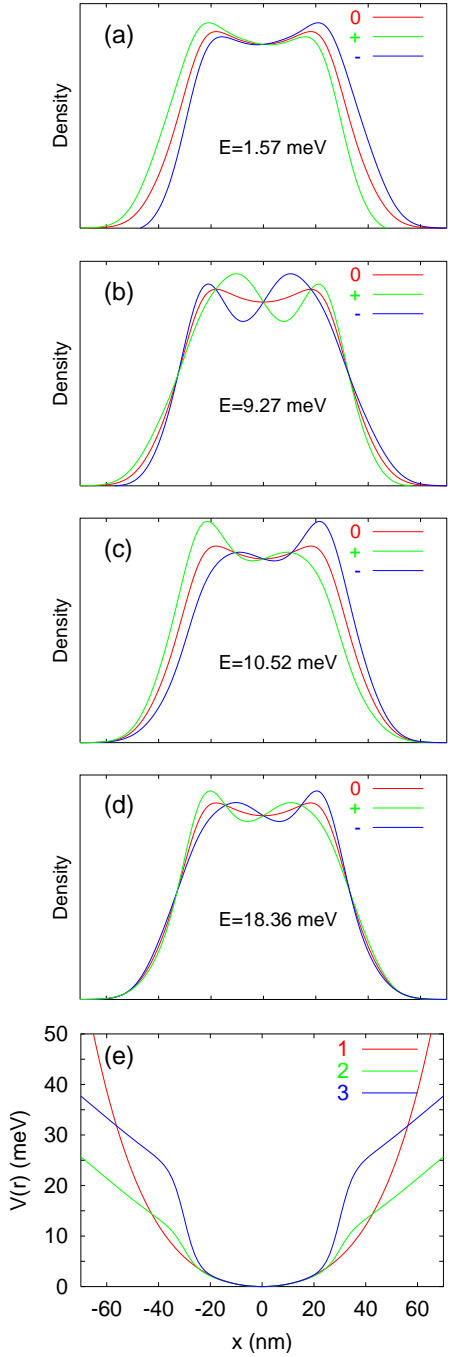}
\end{center}
\caption{Equilibrium density (full curves) and induced density
         (dotted, dashed) for two different moments in time with a phase
         difference of $\pi$. The low-frequency Kohn mode in (a) and the
         high-frequency Kohn mode in (c) represent a nearly perfect
         center-of-mass motion. The below-Kohn mode in (b) and the
         high-frequency `confined' plasmon mode in (d) involve relative
         electron motion with several nodes. (e) shows the different
         potentials used in this paper. Full curve: added $r^4$-term,
         in eq.\ 2: $a=0.48$ meV, $b=14.4\times 10^{-3}$ meV, $c=0$,
         dashed curve: $a=0.48$ meV, $b=-1.8\times 10^{-3}$ meV,
         $c=6$ meV, dotted line: $a=0.48$ meV, $b=-1.8\times 10^{-3}$ meV,
         $c=18$ meV.}
\label{fig5}
\end{figure}

\begin{figure}
\begin{center}
\includegraphics[bb=105 354 440 650,clip,width=8cm]{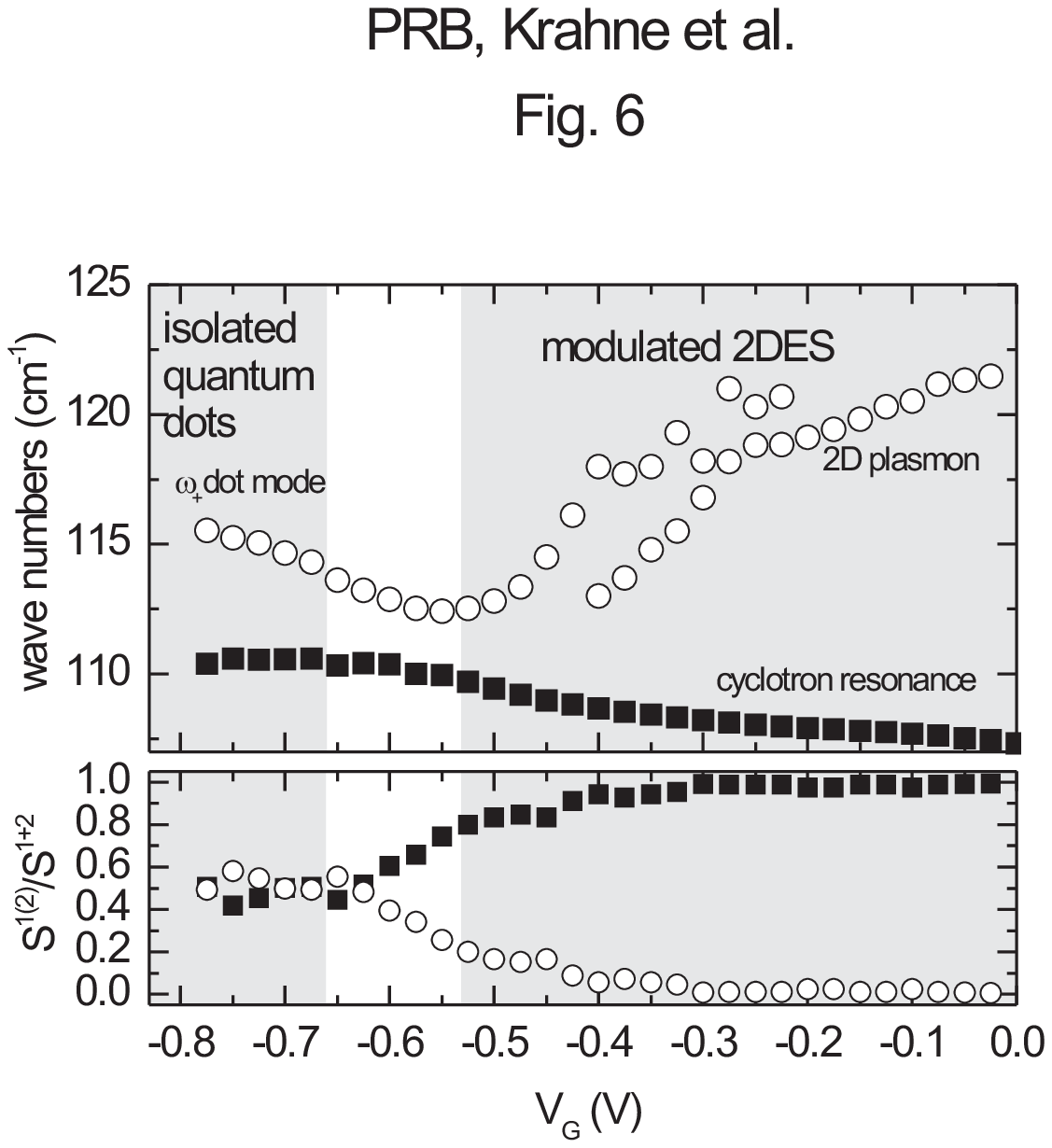}
\end{center}
\caption{Experimental resonance positions at fixed magnetic field
         $B$ = 8 T versus gate voltages. From $V_G$ = 0 to $V_G$ = -0.8 V
         the electron system develops from a density-modulated electron
         system (right shaded regime)  into isolated quantum dots 
         (left shaded regime).
         In the lower panel we plot the relative absorption strengths 
         of the upper
         (empty symbols) and lower mode (full symbols).}
\label{fig6}
\end{figure}

\end{document}